\documentclass[twocolumn,floatfix,amssymb,aps,prb,nobibnotes,nofootinbib]{revtex4-2}
\usepackage{docs}
\usepackage{xcolor}
\usepackage{graphicx}
\usepackage{float}
\usepackage{amsmath}
\usepackage{booktabs}
\usepackage{pifont}
\usepackage[colorlinks=true, linkcolor=blue, citecolor=red, urlcolor=magenta]{hyperref}

\expandafter\ifx\csname package@font\endcsname\relax\else
 \expandafter\expandafter
 \expandafter\usepackage
 \expandafter\expandafter
 \expandafter{\csname package@font\endcsname}%
\fi
\DeclareRobustCommand\substyle{\name@idx{document substyle}}%
\DeclareRobustCommand\classoption{\name@idx{document class option}}%
\DeclareRobustCommand\classname{\name@idx{document class}}%
\def\name@idx#1#2{%
 {\ttfamily#2}%
 \index{#2\space#1=\string\ttt{#2}\space#1}\index{#1>#2=\string\ttt{#2}}%
}%

\begin{document}
\title{Quantum Pontus-Mpemba Effects in Real and Imaginary-time Dynamics}

\author{Hui Yu$^{1}$}
\author{Jiangping Hu$^{1,2}$}
\email{jphu@iphy.ac.cn}
\author{Shi-Xin Zhang$^{1}$}
\email{shixinzhang@iphy.ac.cn}
\affiliation{$^{1}$Beijing National Laboratory for Condensed Matter Physics and Institute of Physics, Chinese Academy of Sciences, Beijing 100190, China}
\affiliation{$^{2}$New Cornerstone Science Laboratory, Institute of Physics, Chinese Academy of Sciences, Beijing 100190, China}

\date{\today}%


\begin{abstract}
The quantum Pontus-Mpemba effect (QPME) is a counterintuitive phenomenon wherein a quantum system relaxes more rapidly through a two-step evolution protocol than through direct evolution under a symmetric Hamiltonian alone. In this protocol, the system first evolves under a symmetry-breaking Hamiltonian and then switches to a symmetric one. We demonstrate that QPME occurs under both real-time and imaginary-time dynamics with respect to $U(1)$-symmetry. Using tilted ferromagnetic initial states, we demonstrate that a transient asymmetric evolution significantly accelerates thermalization or convergence to the ground state for both real-time and imaginary-time evolutions, respectively. The effect is pronounced for small tilt angles, while larger tilts or antiferromagnetic initial states suppress it. Further numerical evidence confirms the consistency of QPME across different system sizes studied. Finally, by implementing a variational optimization procedure, we identify an optimized dynamical path, demonstrating that the relaxation rate can be further enhanced through tailored symmetry-breaking protocols. This work extends the framework of nonequilibrium quantum phenomena to incorporate active state preparation, with direct implications for the implementation of quantum simulation.
\end{abstract}
\maketitle

\section{Introduction}
Non-equilibrium systems exhibit a wealth of exotic phenomena that challenge conventional equilibrium physics, attracting significant scientific attention in recent years. Perhaps the most remarkable of these is the Mpemba effect, a counterintuitive phenomenon in which a hotter system freezes faster than a colder one under identical conditions. First scientifically documented in the 1960s through ice cream freezing experiments \cite{mpemba1969cool}, this effect contradicts classical thermodynamics and has historical roots dating back to ancient \cite{bacon1962opus,descartes1948discours,groves2009now}. Subsequent research has observed this phenomenon across diverse classical systems, including supercooled liquids \cite{esposito2008mpemba,auerbach1995supercooling}, granular fluids \cite{lasanta2017hotter}, clathrate hydrates \cite{ahn2016experimental}, and carbon nanotube resonators \cite{greaney2011mpemba}. Theoretical studies have proposed multiple potential mechanisms: initial state overlap with slow-decaying eigenmodes \cite{lu2017nonequilibrium}, exponentially accelerated relaxation 
\cite{klich2019mpemba,biswas2020mpemba}, and connections to thermal overshooting \cite{antonov2025temperature}. Despite extensive research, a universally accepted explanation for the classical Mpemba effect remains elusive. As a result, this intriguing phenomenon continues to be an active and debated area of study within the scientific community \cite{hu2018conformation, keller2018quenches, holtzman2022landau, baity2019mpemba, kumar2022anomalous, santos2024mpemba, schwarzendahl2022anomalous, teza2023relaxation,pemartin2024shortcuts, gal2020precooling, walker2022mpemba, walker2023optimal, bera2023effect, teza2025speedups}. 

\begin{table*}[ht]
\centering
\resizebox{0.85\textwidth}{!}{ 
\begin{tabular}{c|c|c|c} 
\toprule
& \textbf{Initial states} & \textbf{Hamiltonians} & \textbf{Manifestation} \\
\hline
QME & Different ($\theta_{A}\neq\theta_{B}$) & $H_{sym}$ for both & $O_{A}(t)>O_{B}(t)$ despite $O_{A}(0)<O_{B}(0)$  \\
\hline
QPME & Identical ($\theta_{A}=\theta_{B}$) & $A$: $H_{sym}$ & (1): $O_{B}(t)$ crosses $O_{A}(t)$ before steady state \\ &  & $B$: $H_{asym} \rightarrow H_{sym}$ & (2): $O_{B}(t)$ always below $O_{A}(t)$ \\
\bottomrule
\end{tabular}
}
\caption{Comparison of the QME and the QPME protocols and behavior. Systems $A$ and $B$ evolve under symmetric ($H_{sym}$) or asymmetric ($H_{asym}$) Hamiltonians, with $O(t)$ (or $O(\tau)$ for imaginary-time) denoting the observable. While QME requires different initial states, QPME occurs from identical initial conditions but with a Hamiltonian switch in system $B$.}
\label{table:QME_QPME}
\end{table*}

In recent years, research on the Mpemba effect has expanded beyond classical systems into the quantum realm, where it manifests as the quantum Mpemba effect (QME) \cite{ares2025quantum,yu2025quantum}. The QME can occur through two distinct ways: (1) in open quantum systems coupled to an external environment \cite{nava2019lindblad,carollo2021exponentially,zhang2025observation,moroder2024thermodynamics,aharony2024inverse,kochsiek2022accelerating,wang2024mpemba,boubakour2025dynamical,furtado2024strong,qian2025intrinsic,dong2025quantum,liu2024speeding,ma2025quantum,chatterjee2023quantum,chatterjee2024multiple}, or (2) in isolated quantum systems evolving under unitary dynamics. This work focuses on the latter case. This phenomenon has been observed in both integrable and chaotic quantum systems \cite{ares2023entanglement,liu2024symmetry,turkeshi2025quantum}. The QME manifests when the relaxation curves of interesting physical quantities cross one another for different initial conditions. For example, Refs~\cite{fagotti2014relaxation,essler2016quench,doyon2020lecture,fagotti2014conservation,bertini2015pre,vidmar2016generalized,calabrese2016quantum,polkovnikov2011colloquium,bastianello2022introduction,ares2023lack,summer2025resource} show that $U(1)$-symmetry
restoration occurs more rapidly for more asymmetric initial states under the $U(1)$-symmetric Hamiltonian quench. Likewise, such effect has also been explored in various contexts, including spin chains \cite{khor2024confinement,rylands2024dynamical,murciano2024entanglement,ferro2024non,ares2025simpler,yamashika2025quantum,yu2025tuning,yu2025symmetry,bhore2025quantum,rylands2024microscopic}, many-body localized systems \cite{liu2024quantum}, non-Hermitian systems \cite{guo2025skin,di2025measurement}, Sachdev-Ye-Kitaev models \cite{wang2024mpembaSYK,cao2025symmetry} and random quantum circuits \cite{ares2023entanglement,klobas2025translation,turkeshi2025quantum}. QME also manifests in imaginary time evolution \cite{chang2024imaginary}, where specific initial states of higher energy exhibit faster convergence to the ground state. This anomalous quantum phenomenon has gone beyond theoretical discussion, with experimental verification achieved in various quantum platforms \cite{joshi2024observing,xu2025observation}, marking a significant milestone in non-equilibrium quantum physics.

More recently, the Pontus-Mpemba effect (PME) is proposed \cite{nava2025pontus} which generalizes the standard Mpemba phenomenon through a two-step cooling protocol. In this scheme, two systems initially prepared at the same temperature follow distinct relaxation pathways rather than starting from different initial temperatures. While one system undergoes direct cooling to the target low temperature, the other is first driven to an intermediate state with a higher temperature than the initial state before being quenched to the same final state. The PME occurs when the total relaxation time for the two-step process is shorter than the direct cooling approach. This counterintuitive effect extends the standard Mpemba effect by explicitly incorporating the preparation time of the initial ``hot'' state into full dynamics. Building on these insights, a natural question emerges: Can such non-equilibrium acceleration phenomena occur in closed quantum systems? 

In this work, we identify such non-equilibrium phenomena in both real-time and imaginary-time quantum evolution, with a specific focus on $U(1)$-symmetry. We term this phenomenon as quantum Pontus-Mpemba effect (QPME), distinguishing it from previous studies of the QME. Unlike the QME setup, where two systems $A$ and $B$ are prepared in different $U(1)$-asymmetric initial states and evolve under the same symmetric Hamiltonian $H_{sym}$, the QPME identifies an optimal dynamical pathway for a single initial state via a tailored two-step protocol. Here, systems $A$ and $B$ start in identical $U(1)$-asymmetric initial states. System $A$ evolves continuously under a $U(1)$-symmetric Hamiltonian $H_{sym}$. System $B$, however, first evolves under a $U(1)$-asymmetric Hamiltonian $H_{asym}$ for a controlled time interval before undergoing an abrupt transition to $H_{sym}$. Remarkably, the QPME manifests in two ways: (1) The time evolution of a given physical observable in system $B$ intersects that of system $A$ before both reach steady state, with $B$ exhibiting faster relaxation after the crossing, or (2) the evolution curve of system $B$ remains entirely below that of system $A$ at all times. Both cases demonstrate that the two-step protocol leads to accelerated relaxation compared to purely symmetric evolution. A comparison of the QME and QPME is presented in Table~\ref{table:QME_QPME}. This operational distinction makes the QPME a more practical and active strategy for accelerating quantum state preparation and probing non-equilibrium dynamics in settings where one seeks an efficient path from a given initial condition to a target state, rather than comparing different starting points.

The remainder of this work is structured as follows. Section II introduces our theoretical framework, including the system setup and key observables. In Section III, we present comprehensive numerical results for both real-time and imaginary-time evolution, with a detailed analysis of the observed phenomena. In Section IV, we employ a variational optimization approach to identify an optimized dynamical path towards equilibrium. Finally, Section V provides concluding remarks and discusses potential implications of our findings.

\section{Setup}
We begin by emphasizing the key distinctions between real-time and imaginary-time evolution. Real-time evolution describes the unitary dynamics of a quantum system, where energy is strictly conserved and the system thermalizes for general chaotic evolution, effectively restoring symmetry within subsystems in the long-time limit. In contrast, imaginary-time evolution is a non-unitary and dissipative process that projects an initial state onto the ground state by exponentially suppressing higher-energy components. Unlike real-time dynamics, it does not conserve energy and symmetry charges even with a symmetric Hamiltonian, fails thermalization, and does not restore symmetries. Instead, it systematically drives the system toward the lowest energy ground state.

Our study focuses on two families of initial states: the ferromagnetic state $\vert 000...0 \rangle$, the antiferromagnetic N\'{e}el state $\vert 0101..1 \rangle$. To incorporate the effect of symmetry breaking in the initial state, we introduce tilted ferromagnetic states, defined as 
\begin{eqnarray}
    \vert \psi_{i} (\theta)\rangle =  e^{-i\frac{\theta}{2} \sum_{j} \sigma_{j}^{y}} \vert 000...0 \rangle
    \label{eq: tilted ferromagnetic}
\end{eqnarray}
Where $\theta$ parametrizes the degree of symmetry breaking in the initial state. At $\theta=0$, the state preserves $U(1)$-symmetry. Tilted antiferromagnetic states are constructed analogously by applying the same rotation to the Néel state. In this work, we study both real-time and imaginary-time dynamics. Under real-time evolution, the wavefunction at time $t$, $|\psi(t)\rangle$, evolves unitarily as $e^{-i\hat{H}t}|\psi_{i} (\theta)\rangle$. In contrast, imaginary-time evolution \cite{alipour2025state} yields:
\begin{eqnarray}
|\psi(\tau)\rangle=\frac{e^{-\hat{H}\tau}|\psi_{i}(\theta)\rangle}{\sqrt{\langle \psi_{i}(\theta)|e^{-2\hat{H}\tau}|\psi_{i}(\theta)\rangle}}
\end{eqnarray}
where the denominator ensures proper normalization and $\tau$ denotes imaginary time. 

We investigate the disordered one-dimensional spin chain Hamiltonian $H$ with broken $U(1)$-symmetry and periodic boundary conditions:
\begin{eqnarray}
H = & - \sum_{j=1}^{L}(\sigma_j^x \sigma_{j+1}^x + \gamma \sigma_j^y \sigma_{j+1}^y + \mu \sigma_j^z \sigma_{j+1}^z) \notag \\
& + \sum_{j=1}^{L} h_{j}\sigma_j^z. 
\label{eq:Ham2}
\end{eqnarray}
where $L$ is the total system size and $\sigma_{j}^{\alpha}$ ($\alpha=x,y,z$) are Pauli matrices acting on site $j$. The periodic boundary conditions enforce $\sigma_{L+1}^{\alpha}=\sigma_{1}^{\alpha}$  for $\alpha=x,y,z$. $\mu=-0.5$ sets the nearest-neighbor interaction strengths. Disorder is introduced via random fields $h_{j}$ along the $z$-axis, sampled uniformly from $[-W, W]$ with $W=1$. In the Supplemental Material (SM), we calculate the level spacing ratio \cite{oganesyan2007localization} within the half-filling sector to ensure that the Hamiltonian $H$ is chaotic in that parameter regime. The parameter $\gamma$ serves as a control for $U(1)$-symmetry breaking in the system: at $\gamma=1$, the Hamiltonian preserves $U(1)$-symmetry and is denoted $H_{sym}$,  while any deviation from unity explicitly breaks this symmetry, yielding the asymmetric Hamiltonian $H_{asym}$. 

\begin{figure}[htbp]
\begin{center}
\includegraphics[scale=0.435]{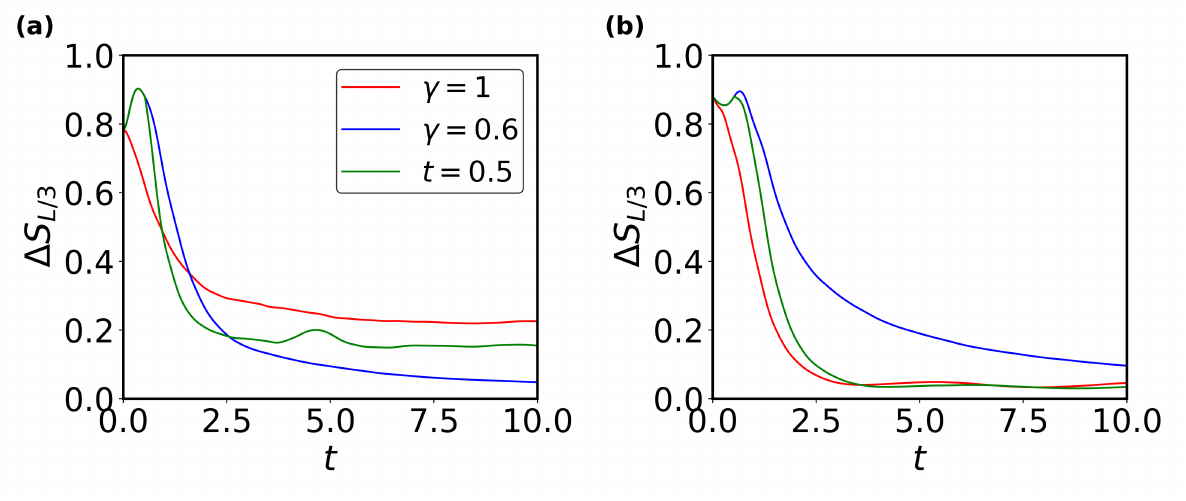}
\caption{The real-time dynamics of the entanglement asymmetry $\Delta S_{L/3}$ is illustrated for two distinct initial states with a tilted angle $\theta=0.2\pi$. (a) the tilted ferromagnetic state and (b) the tilted antiferromagnetic state. The red and blue curves represent symmetric and asymmetric evolution, respectively, where the asymmetric evolution is governed by a Hamiltonian with $\gamma=0.6$. Additionally, the green curves illustrate a two-step protocol with a switch time set to $t=0.5$.}
\label{fig:EA_real_time}
\end{center}
\end{figure}

To investigate relaxation dynamics in this model, we analyze three key observables: the entanglement asymmetry, the energy expectation, and the charge variance. 

Entanglement asymmetry (EA) has been utilized extensively as a powerful tool for characterizing symmetry breaking in both quantum field theories \cite{capizzi2024universal,capizzi2023entanglement,chen2024renyi} and quantum many-body systems \cite{rylands2024microscopic,khor2024confinement,ares2025entanglement}. To define entanglement asymmetry, we typically partition the entire system into a subsystem $A$ and its complement $\overline{A}$. The reduced density matrix for subsystem $A$, denoted as $\rho_{A}$, is obtained by tracing out the degrees of freedom in $\overline{A}$, expressed as $\rho_{A}=\text{Tr}_{\overline{A}}(\rho)$. Here, $\rho$ represents the full-system density matrix. For real-time evolution, $\rho=|\psi(t)\rangle\langle\psi(t)|$, while for imaginary time, it is normalized as $\rho=|\psi(\tau)\rangle\langle\psi(\tau)|$. The entanglement asymmetry is defined as:
\begin{eqnarray}
    \Delta S_{A}  = S(\rho_{A, Q}) - S(\rho_{A}),
\end{eqnarray}
where $S(\rho)$ is the von Neumann entanglement entropy, given by $S(\rho)=-\text{Tr}(\rho\log\rho)$, and $\rho_{A, Q}$ is the symmetry-resolved reduced density matrix:
\begin{eqnarray}
\rho_{A, Q} = \sum_{q \in \mathbb{Z}} \Pi_{q} \rho_{A} \Pi_{q}
\end{eqnarray}
Here, $\hat{Q}_{A} = \sum_{i \in A} \sigma_{i}^{z}$ is the subsystem charge operator for $U(1)$-symmetry and $\Pi_{q}$ projects onto the eigenspace of $\hat{Q}_{A}$ with charge $q$. Consequently, $\rho_{A, Q}$ is block diagonal in the eigenbasis of $\hat{Q}_{A}$. The EA satisfies two key properties: (1) $\Delta S_{A} \geq 0$ since the EA is defined as the relative entropy between $\rho_{A, Q}$ and $\rho_{A}$. (2) $\Delta S_{A} = 0$ if and only if $\rho_{A, Q}=\rho_{A}$. The EA measures the distinguishability between the reduced density matrix $\rho_{A}$ and its symmetry-resolved counterpart $\rho_{A, Q}$, providing a quantitative diagnostic for symmetry breaking in $A$.

In imaginary-time evolution, the expectation value of energy is the object of interest. For the state $|\psi(\tau)\rangle$, the energy expectation is given by:
\begin{eqnarray}
    \langle \hat{H} \rangle_{\tau} = \langle \psi(\tau)|\hat{H}|\psi(\tau)\rangle=\frac{\langle \psi_{i}(\theta)|e^{-\hat{H}\tau}\hat{H}e^{-\hat{H}\tau}|\psi_{i}(\theta)\rangle}{\langle \psi_{i}(\theta)|e^{-2\hat{H}\tau}|\psi_{i}(\theta)\rangle} \label{H_exp} 
\end{eqnarray}
Expanding the initial state $|\psi_{i}(\theta)\rangle$ in the eigenbasis $\{ |n\rangle \}$ of $\hat{H}$ with eigenvalues $E_{n}$, we obtain:
\begin{eqnarray}
    \langle \hat{H} \rangle_{\tau} = \frac{\sum_{n}|c_{n}|^{2}e^{-2E_{n}\tau}E_{n}}{\sum_{n}|c_{n}|^{2}e^{-2E_{n}\tau}} \label{H_exp_expansion}
\end{eqnarray}
where $c_{n}=\langle n|\psi_{i}(\theta)\rangle$ are the overlap coefficients for the initial states. While the energy expectation value remains conserved under real-time dynamics,  $\langle \hat{H} \rangle_{\tau}$ monotonically decreases to the ground state energy $E_{0}$ as $\tau$ goes to infinity when higher-energy terms ($E_{n}>E_{0}$) are exponentially suppressed. In imaginary-time evolution, our target is the ground state of $H_{sym}$.

\begin{figure}[htbp]
\begin{center}
\includegraphics[scale=0.485]{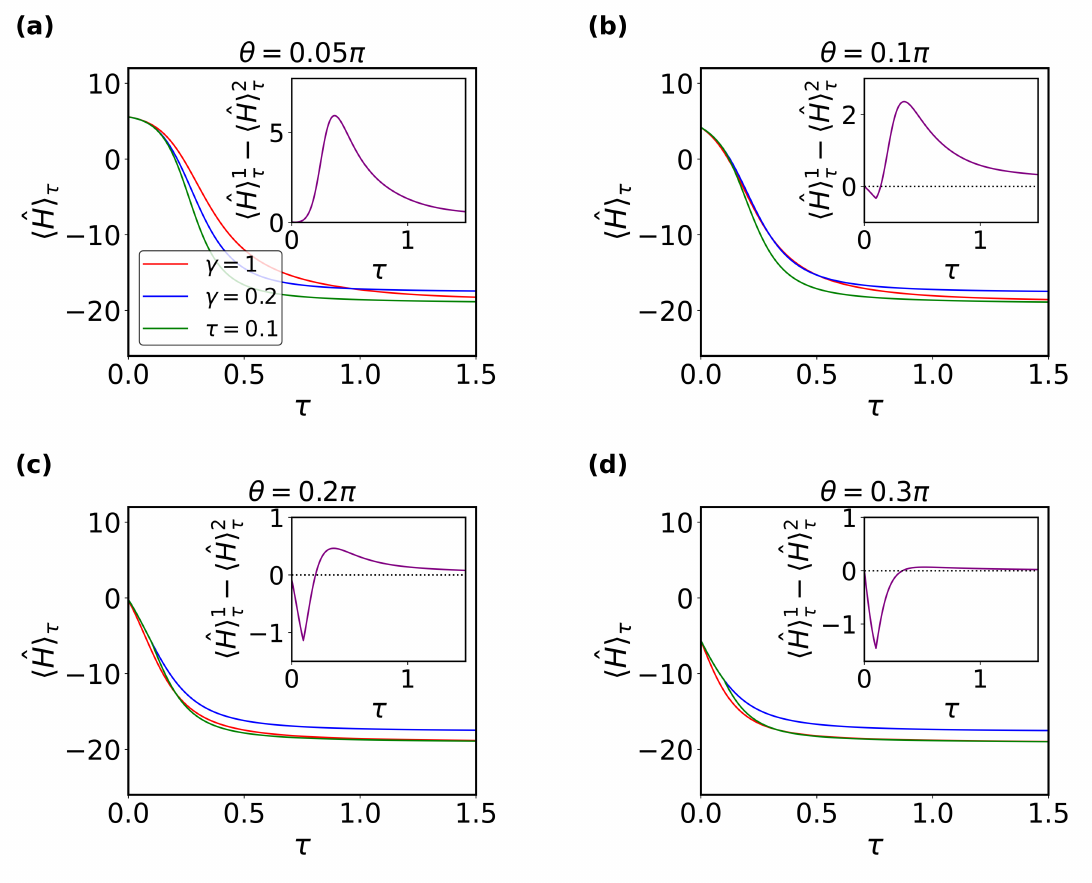}
\caption{The imaginary-time evolution of $\langle \hat{H} \rangle_{\tau}$ is presented for tilted ferromagnetic states across different tilted angles: (a) $\theta=0.05\pi$, (b) $\theta=0.1\pi$, (c) $\theta=0.2\pi$, and (d) $\theta=0.3\pi$. The red and blue curves correspond to symmetric and asymmetric evolution, respectively, with all asymmetric evolution governed by the Hamiltonian with $\gamma=0.2$. The green curves depict results from a two-step protocol where the evolution switches from asymmetric to symmetric at $\tau=0.1$. The insets display the energy difference between the direct quench $\langle \hat{H} \rangle_{\tau}^{1}$, and the two-step protocol $\langle \hat{H} \rangle_{\tau}^{2}$ as a function of $\tau$. A positive (negative) difference indicates that the energy expectation value of the two-step protocol relaxes faster (slower) than that of the direct quench.}
\label{fig:E_imag_time_F}
\end{center}
\end{figure}

Finally, we investigate the behavior of charge variance (CV) during imaginary-time evolution. The CV is defined as:
\begin{eqnarray}
\sigma^{2}_{Q}= \langle \hat{Q}^{2} \rangle - \langle \hat{Q} \rangle^{2}
\end{eqnarray}
where $\hat{Q} = \sum_{i=1}^{L} \sigma_{i}^{z}$ is the total charge operator. The expectation values $\langle \hat{Q}\rangle$ and $\langle \hat{Q^{2}}\rangle$ are computed analogously to Eq.~\eqref{H_exp}, with $\hat{H}$ replaced with $\hat{Q}$ or $\hat{Q}^{2}$, respectively. If the expectation value $\langle \cdot \rangle$ is taken with respect to a $U(1)$-symmetric state where the state resides entirely within a single charge sector, then $\sigma^{2}_{Q}=0$. Conversely, a nonzero CV indicates that the state spans multiple charge sectors, reflecting symmetry breaking. The CV thus serves as a complementary measure to entanglement asymmetry for probing relaxation dynamics, directly quantifying fluctuations in the charge distribution. 

Furthermore, all observables require statistical averaging over an ensemble of disorder realizations for the Hamiltonian to obtain reliable results.

\begin{figure}[htbp]
\begin{center}
\includegraphics[scale=0.435]{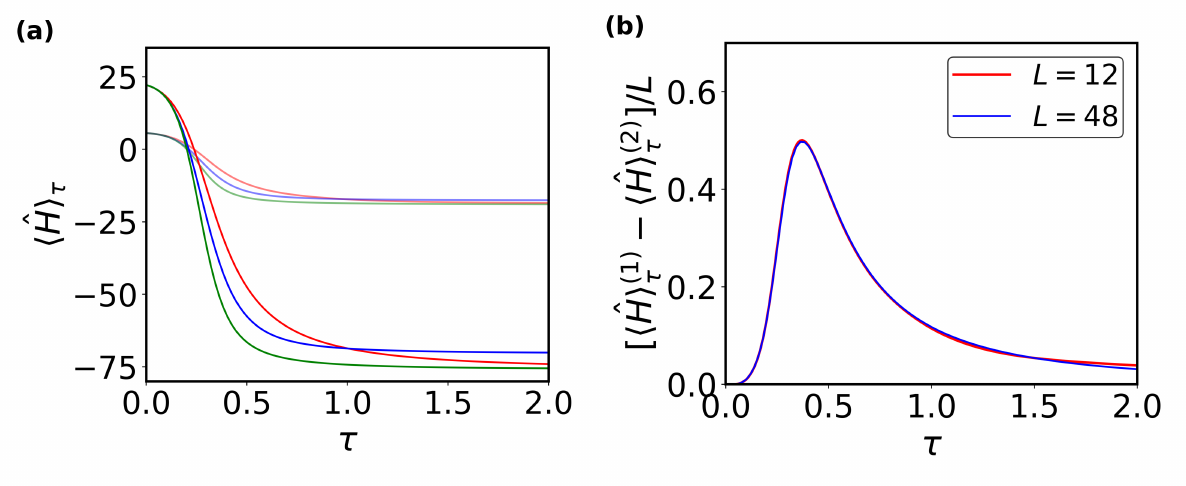}
\caption{(a) Imaginary-time evolution of $\langle \hat{H} \rangle_{\tau}$ for tilted ferromagnetic states ($\theta=0.05\pi$) at system sizes $L=12$ (light colors) and $L=48$ (dark colors), with red, blue, and green curves maintaining have the same meaning as Fig.~\ref{fig:E_imag_time_F}.  (b) Energy density difference between two protocols as a function of $\tau$ for both $L=12$ and $L=48$. $\langle \hat{H} \rangle_{\tau}^{1}$ and $\langle \hat{H} \rangle_{\tau}^{2}$ represent the energy expectation values calculated from direct quench and two-step protocol evolution, respectively.}
\label{fig:E_imag_time_L}
\end{center}
\end{figure}

\section{Results}
All numerical simulations are performed using the {\sf TensorCircuit-NG} package \cite{zhang2023tensorcircuit}, with results obtained through ensemble averaging over 500 independent disorder realizations.  We set the system size to $L=12$ for most of our analysis, but expand to $L=48$ when investigating finite-size effects and the scalable robustness of the QPME.

\subsection{Real-time Dynamics}
We investigate the dynamics of entanglement asymmetry by considering a subsystem $A$ of 4 sites and setting the switching time from $H_{asym}$ to $H_{sym}$ at $t=0.5$. Our analysis compares a direct relaxation under $H_{sym}$ with our two-step protocol, revealing distinct behaviors for different initial states. For a tilted ferromagnetic initial state with a tilt angle of $\theta=0.2\pi$, we observe the clear emergence of case (1) of QPME, as shown in Fig.~\ref{fig:EA_real_time}(a). This is identified by the crossing of the EA curves for the direct quench (red) and the two-step protocol (green) around $t=1$. The underlying mechanism is the phenomenon of Hilbert subspace imprint (HSI) \cite{yu2025hilbert}. This phenomenon occurs when a system, initially prepared in a specific symmetry sector of a symmetric Hamiltonian $H_{0}$, is subjected to a weak symmetry breaking perturbation $H_{1}$. In this scenario, the initial state's overlap remains predominantly confined to a polynomially small set of eigenstates of the perturbed Hamiltonian $H=H_{0}+H_{1}$, as these eigenstates maintain an adiabatic connection to those of $H_{0}$. This dynamical confinement thereby leads to non-thermalizing behavior. To understand why the direct quench is slow, we note that the dynamics of the tilted ferromagnetic state under $H_{sym}$ are unitarily equivalent to the dynamics of the ferromagnetic state under a different, corresponding weak asymmetric Hamiltonian $H'$ (see SM for details). In this equivalent picture, the symmetry-breaking terms in $H'$ are proportional to $\sin{\theta}$ and act as a weak perturbation for small tilt angles. According to HSI, such weak symmetry breaking approximately confines the dynamics of the ferromagnetic state to a polynomially small corner of the Hilbert space, causing slow, non-thermal relaxation. The two-step protocol overcomes this limitation as the initial evolution under the symmetry-breaking Hamiltonian enhances the thermal properties of the state by driving the state into a broader charge distribution and promoting its spread across the Hilbert space. This process destroys the condition for HSI and accelerates the subsequent relaxation when the system is switched to $H_{sym}$, leading to the observed QPME. We further observe that the green and red curves converge to different steady-state values, despite evolving under the same final Hamiltonian after the switch. This difference arises because direct evolution under $H_{sym}$ alone results in a non-thermal steady state due to HSI, whereas the two-step protocol drives the system toward a thermal state. However, the QPME vanishes for larger tilt angles, such as $\theta=0.4\pi$ (see SM). At a large tilt, the initial state is already distributed broadly across different charge sectors with no remnants of HSI. This leaves no room for the transient asymmetric evolution to further enhance thermalization, thereby eliminating the speed-up effect. Furthermore, we note that the transient asymmetric evolution induces a characteristic overshooting in the EA (green curve), a feature that has been previously reported in \cite{yu2025symmetry} for symmetry-breaking dynamics.

In stark contrast, QPME is absent for the tilted antiferromagnetic initial states, as demonstrated for angles $\theta=0.2\pi$ and $\theta=0.4\pi$ in Fig.~\ref{fig:EA_real_time}(b) and SM. Here, the EA for the two-step protocol remains consistently above that of the direct quench.This absence stems from the intrinsic properties of the antiferromagnetic state: it initially possesses a high overlap with an exponential number of eigenstates within the vast $Q=0$ sector. It is therefore inherently ``thermal'', meaning it efficiently explores the available state space and relaxes rapidly under $H_{\text{sym}}$ alone, without suffering from the HSI confinement that plagues the ferromagnetic case. Consequently, the initial asymmetric evolution provides no additional thermalization advantage, and no QPME is observed. As the initial state is already ``thermal'' in nature, both protocols ultimately converge to the same steady-state EA value.

\begin{figure}[htbp]
\begin{center}
\includegraphics[scale=0.435]{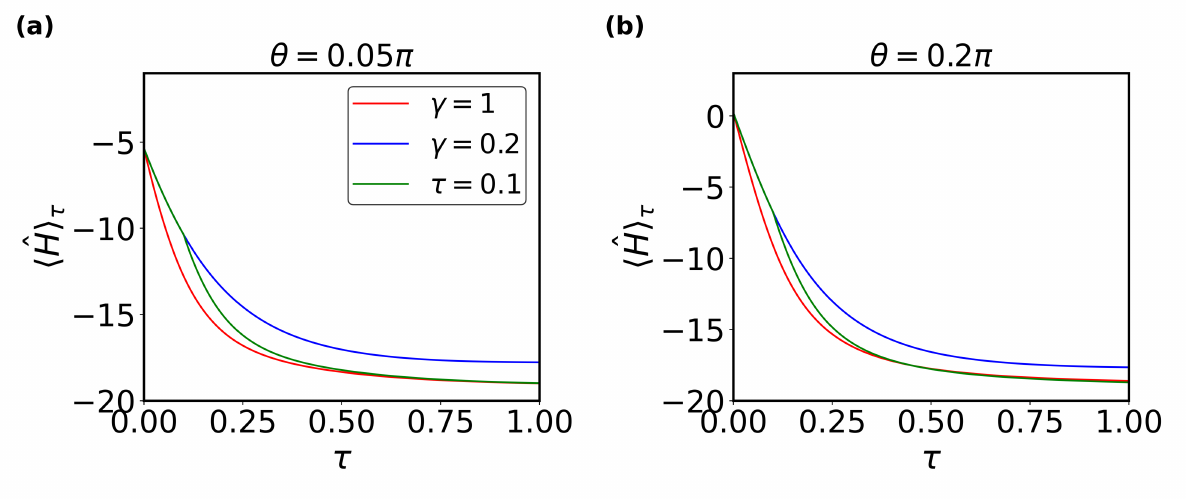}
\caption{The imaginary-time dependence of $\langle \hat{H} \rangle_{\tau}$ is shown for tilted antiferromagnetic states at two different angles: (a) $\theta=0.05\pi$, and (b) $\theta=0.2\pi$. The red and blue curves correspond to symmetric and asymmetric (with $\gamma=0.2$) evolution, respectively, while green curves show the two-step protocol where the Hamiltonian switches from asymmetric to symmetric at $\tau=0.1$.}
\label{fig:E_imag_time_AF}
\end{center}
\end{figure}

\subsection{Imaginary-time Dynamics}
In this section, we analyze the energy dynamics during imaginary-time evolution for both protocols, using a switching time of $\tau=0.1$. The primary objective is to identify protocols that enable faster convergence to the ground state in imaginary-time evolution, which is both helpful in numerical algorithms and experiment protocols. 

As illustrated in Fig.~\ref{fig:E_imag_time_F}(a) and (b), at small tilt angles $\theta$, the two-step protocol yields a more rapid convergence to the ground state compared to the direct quench for tilted ferromagnetic states. We note that the ground state of $H_{sym}$ resides in the $Q=0$ sector for most disorder configurations (see SM). Specifically, in Fig.~\ref{fig:E_imag_time_F}(a), the energy difference between the two protocols remains consistently positive throughout the evolution. This observation, detailed in the inset, corresponds to case (2) of QPME. As the tilt angle is slightly increased to $\theta=0.1\pi$, a qualitative change in the energy dynamics occurs, as shown in Fig.~\ref{fig:E_imag_time_F}(b). The inset reveals that the energy difference between the protocols changes sign during the evolution, indicating a transition to case (1) of QPME. However, as the tilt angle increases further, the QPME diminishes and eventually vanishes, as evidenced in Fig.~\ref{fig:E_imag_time_F}(c) and (d). This accelerated imaginary-time relaxation arises because the initial asymmetric evolution phase in the two-step protocol drives the state toward a broader charge distribution by the switching time. Although imaginary-time evolution under $H_{sym}$ also breaks $U(1)$ symmetry, the stronger symmetry breaking induced by $H_{asym}$ leads to more rapid charge redistribution at the initial stage. This enhanced charge broadening allows the state to relax more efficiently after the switch to $H_{sym}$ compared to the relatively narrower distribution produced by direct quench (see SM).  As the tilt angle increases, the initial state already possesses a broader charge distribution, reducing the relative impact of the transient asymmetric evolution and thus suppressing the QPME. This transition can also be understood from the perspective of eigenstate decomposition. States that relax more rapidly exhibit a higher initial overlap with higher-energy eigenstates of the asymmetric Hamiltonian $H_{asym}$, since these components decay faster during imaginary-time evolution \cite{chang2024imaginary}. This phenomenon is most pronounced at small tilt angles. As shown in the SM, the tilted ferromagnetic state with $\theta = 0.05\pi$ has greater overlap with higher-energy eigenstates of $H_{\text{asym}}$ than with those of $H_{\text{sym}}$, further elucidating the mechanism behind the accelerated relaxation in the two-step protocol. Despite these $\theta$-dependent differences in the relaxation dynamics, all green and red curves in Fig.~\ref{fig:E_imag_time_F} ultimately converge to the same final value: the ground state energy of $H_{sym}$. 

To investigate the persistence of the QPME, we compare energy dynamics for a system of size $L=12$, computed via exact methods, with a significantly large system of $L=48$, simulated using time-evolving block decimation (TEBD) \cite{vidal2004efficient}. As expected for an extensive quantity, the total energy in the $L=48$ system is four times larger than in the smaller system (Fig.~\ref{fig:E_imag_time_L}(a)). To directly quantify the QPME, we analyze the energy density difference between the two protocols. The result, presented in Fig.~\ref{fig:E_imag_time_L}(b), demonstrates that the difference curves for $L=12$ and $L=48$ are nearly identical. This observed consistency across system sizes indicates that the relative speed-up from the two-step protocol is not a finite-size artifact and suggests its manifestation in larger systems. We also investigate the effect of different switching times and find that the QPME is most pronounced when the switching time is significantly shorter than the system's intrinsic relaxation timescale (see SM).

\begin{figure}[htbp]
\begin{center}
\includegraphics[scale=0.485]{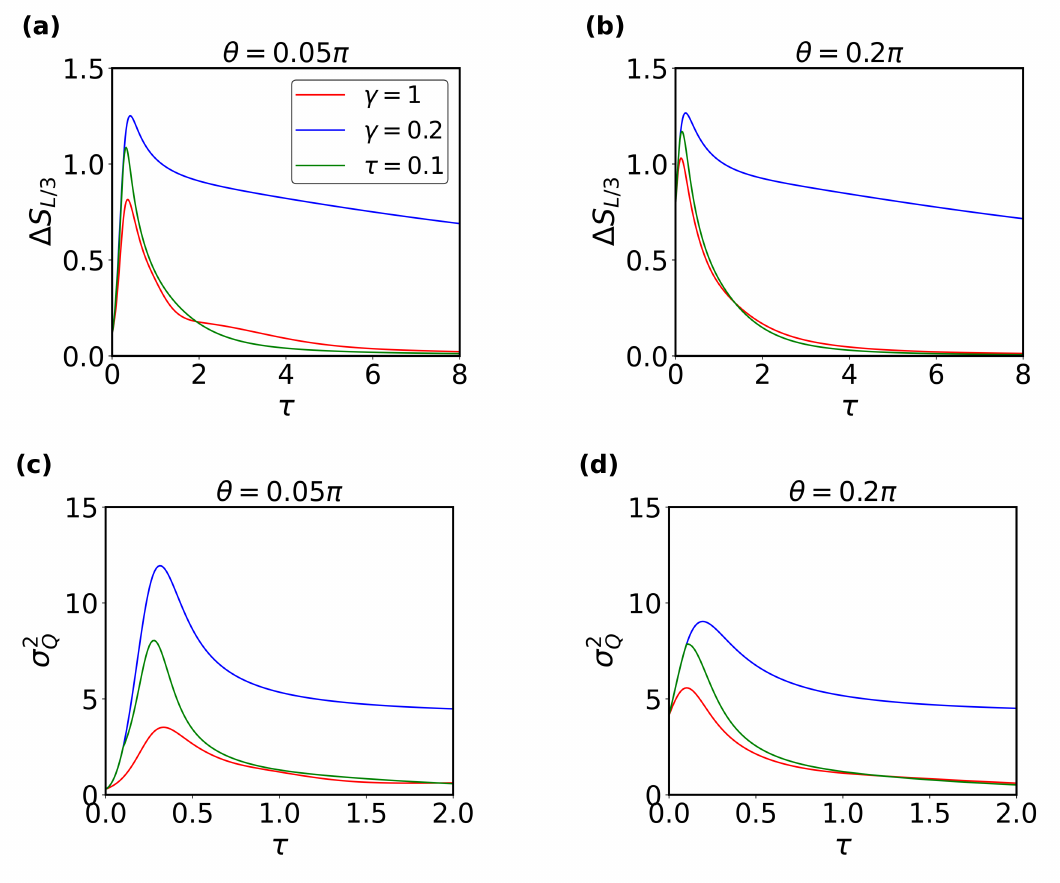}
\caption{The evolution of the entanglement asymmetry, $\Delta S_{L/3}$, is presented as a function of $\tau$ for tilted ferromagnetic states at two different angles: (a) $\theta=0.05\pi$, and (b) $\theta=0.2\pi$. In panels (c) and (d), the imaginary-time evolution of the charge variance, $\sigma_{Q}^{2}$, is examined for the same tilted ferromagnetic states at the corresponding angles. In all panels, the red curves represent the symmetric evolution, while the blue curves illustrate the asymmetric case with $\gamma=0.2$. Additionally, the green curves display the two-step protocol, where the dynamics switch to a symmetric evolution at time $\tau=0.1$.}
\label{fig:EAandCV_imag_time_F}
\end{center}
\end{figure}

In contrast to the ferromagnetic case, tilted antiferromagnetic initial states exhibit no QPME. As shown in Fig.~\ref{fig:E_imag_time_AF}(a) and (b), the two-step protocol provides no speed-up for any of the tested tilt angles. This absence of QPME can be attributed to the symmetry sectors of the initial and final states. Both the initial antiferromagnetic state and the target ground state of $H_{sym}$ reside in the same $Q=0$ charge sector. Consequently, the initial period of asymmetric evolution is ineffective at redistributing probability weight across different charge sectors, thereby eliminating any potential for the QPME. To complement our analytical understanding, we study the dynamics of entanglement asymmetry and charge variance in imaginary time using the same parameters as in Fig.~\ref{fig:E_imag_time_F}(a) and (c). Fig.~\ref{fig:EAandCV_imag_time_F} shows that the transient asymmetric evolution actively redistributes charge, evidenced by the larger EA and CV in the two-step protocol versus direct quench. Although evolution under $H_{sym}$ also alters the initial charge configuration, this process occurs more slowly than under $H_{asym}$. Early, both red and green curves in Fig.~\ref{fig:EAandCV_imag_time_F}(a) and (b) exhibit an overshooting as seen in the green curves of Fig.~\ref{fig:EA_real_time}. Ultimately, both EA and CV converge to the same late-time values, consistent with the behavior observed in energy evolution. 

To conclude, the initial period of asymmetric evolution serves a unified purpose in both real- and imaginary-time dynamics: broadening the charge distribution of the initial state. However, the consequence of this broadening differs fundamentally between the two contexts. In real-time dynamics, a broader distribution drives the system toward a thermal state away from HSI restriction, thereby enhancing subsequent relaxation under the symmetric Hamiltonian. In imaginary-time dynamics, the broader distribution accelerates convergence to the ground state residing in the half-filling sector. 

\section{Variational optimization of the dynamical path}
\begin{figure}[htbp]
\begin{center}
\includegraphics[scale=0.29]{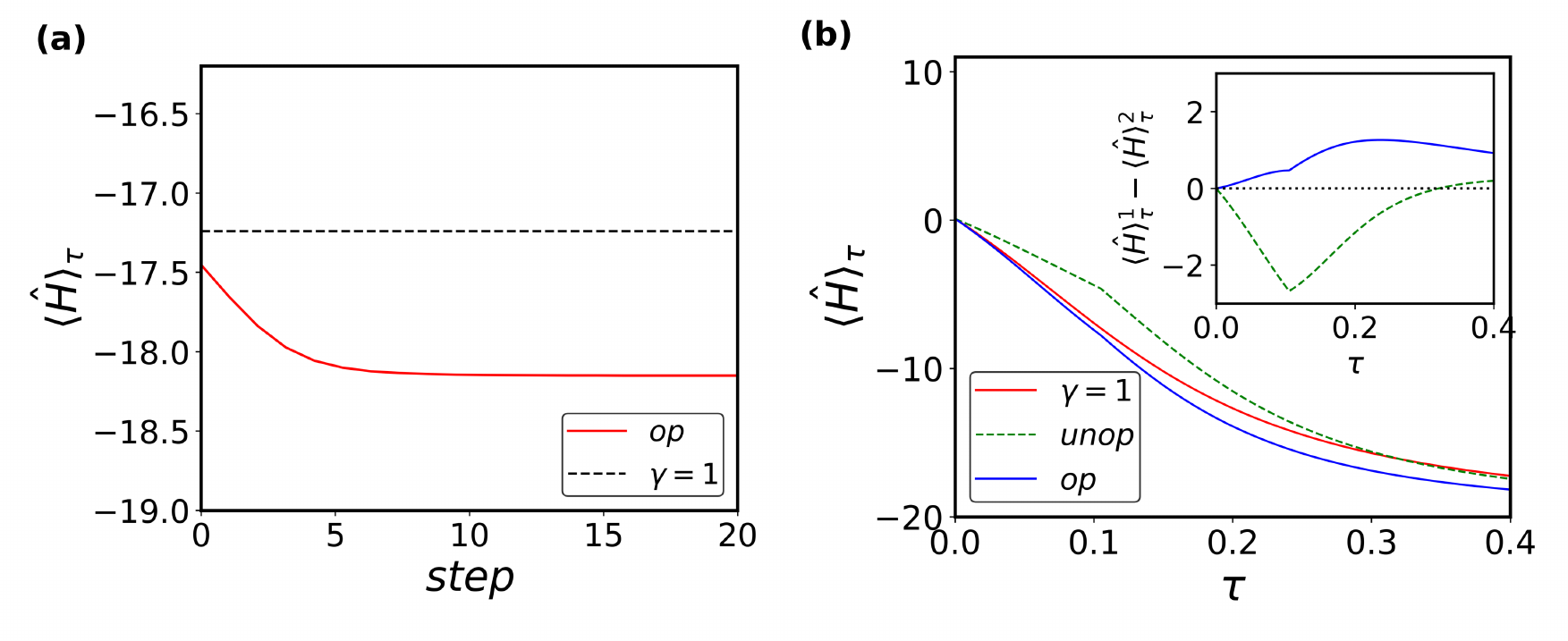}
\caption{Variational identification of the relaxation path for $L=12$ and tilted ferromagentic state with $\theta=0.2\pi$. (a) Convergence of the loss function (energy at $\tau_{total}=0.4$) over optimization steps. The switching time is fixed at $\tau=0.1$. The black dashed line indicates the reference energy for a system governed solely by $H_{sym}$. (b) Energy relaxation during imaginary-time evolution. The red line shows the standard direct quench under $H_{sym}$. The two-step protocols involve an initial evolution under $H_{opt}$ followed by $H_{sym}$. The blue line utilizes the optimized parameters $\{\gamma_{j},h_{j}\}$, while the green dashed line uses unoptimized parameters. The inset displays the energy difference between the direct quench $\langle \hat{H} \rangle_{\tau}^{1}$, and the two-step protocol $\langle \hat{H} \rangle_{\tau}^{2}$ as a function of $\tau$. A positive (negative) difference indicates that the energy expectation value of the two-step protocol relaxes faster (slower) than that of the direct quench.}
\label{fig:optimal}
\end{center}
\end{figure}

In the previous sections, we demonstrated the QPME using a uniform symmetry-breaking Hamiltonian $H_{asym}$ with a fixed parameter $\gamma$. A natural question arises: can we identify a theoretically optimized path that enhances the relaxation rate? To rigorously identify a highly efficient dynamical path towards equilibrium without relying on an arbitrary ansatz, we implement a variational optimization procedure over the parameters of the transient Hamiltonian. We parameterize the transient Hamiltonian $H_{opt}$ using the same form as Eq.~\eqref{eq:Ham2} but relax the constraints to allow for site-dependent couplings and fields. This yields a high-dimensional variational landscape defined by:
\begin{eqnarray}
H_{opt}(\{\gamma_{j}\},\{h_{j}\}) = & - \sum_{j=1}^{L}(\sigma_j^x \sigma_{j+1}^x  + \mu \sigma_j^z \sigma_{j+1}^z) \notag \\
& - \sum_{j=1}^{L}\gamma_{j}\sigma_j^y \sigma_{j+1}^y + \sum_{j=1}^{L} h_{j}\sigma_j^z.
\end{eqnarray}
Here, the variational parameters are the bond-dependent anisotropy $\gamma_{j} \in[0,1]$, which controls the strength of the symmetry-breaking term on each bond, and the site-dependent longitudinal fields $h_{j} \in [-1,1]$. This parameterization covers a wide class of local symmetry-breaking Hamiltonians. 

We define the efficiency of the dynamical path by the proximity of the system to the target ground state after a fixed total evolution time, $\tau_{total}$. The optimization objective is to minimize the expectation value of the energy with respect to the target symmetric Hamiltonian $H_{sym}$ (where $\gamma_{j}=1$ for all $j$):
\begin{eqnarray}
\mathcal{L}  = \langle \psi(\tau_{total})|\hat{H}_{sym}|\psi(\tau_{total})\rangle
\end{eqnarray}
where $|\psi(\tau_{total})\rangle$ is the state obtained by evolving the initial state first under $H_{opt}(\{\gamma_{j}\},\{h_{j}\}) $ for time $\tau$, and subsequently under $H_{sym}$ for the remainder of the time. In the optimization procedure, we enforce physical constraints on the variational parameters to ensure that $\gamma_{j} \in[0,1]$ and $h_{j} \in [-1,1]$ throughout the search. This is implemented via a projected gradient-descent method, where after each update step, any parameter exceeding its prescribed bounds is clipped back to the nearest boundary. Such regularization guarantees that the transient Hamiltonian remains physically admissible and allows for a meaningful comparison with the unoptimized, uniform perturbation case. To compare with the unoptimized results from Fig.~\ref{fig:E_imag_time_F}(c), we perform this optimization for the imaginary-time dynamics of a tilted ferromagnetic state ($\theta=0.2\pi$) with a system size $L=12$. We fix the total evolution time $\tau_{total}=0.4$ and the switching time $\tau=0.1$. The optimizer is initialized with uniform couplings $\gamma_{j}=0.2$ for all $j$ and random longitudinal fields, using a learning rate of $0.4$. 

As shown in Fig.~\ref{fig:optimal}(a), the loss function (final energy) decreases monotonically as the optimization progresses and saturates after approximately 8 steps. This indicates that the gradient descent algorithm consistently converges to a specific, non-trivial configuration of couplings $\{\gamma_{j}\}$ and fields $\{h_{j}\}$. To verify that this configuration indeed yields a faster relaxation rate, we analyze the time dependence of the energy in Fig.~\ref{fig:optimal}(b). For the two-step quench using unoptimized parameters (green dashed line), we observe the first scenario of QMPE described in Table.~\ref{table:QME_QPME}. This is characterized by a crossing between the two-step evolution and the direct quench (red line) around $\tau=0.32$, implying the advantage only emerges at later times. In contrast, when employing the optimized parameters for $H_{opt}$ (blue line), we observe another case of QMPE behavior, where the energy remains strictly below the direct quench baseline throughout the entire evolution. This persistent advantage is quantified in the inset, where the energy difference between the direct quench and the optimized protocol remains positive for all $\tau$. 

We also extend this variational approach to real-time dynamics to minimize the entanglement asymmetry, with detailed results provided in the SM. In doing so, it is important to address a subtle distinction regarding the definition of an optimized paths. In imaginary-time evolution, the system universally converges to the unique ground state of $H_{sym}$, ensuring an identical target for both direct and two-step protocols. However, in real-time dynamics, the system thermalizes to a steady state determined by its energy density. Since the initial evolution under the variational Hamiltonian $H_{opt}$ modifies the system's energy relative to $H_{sym}$, the two-step protocol ultimately targets a thermal ensemble with a different effective temperature than the direct quench. Nevertheless, the entanglement asymmetry eventually reaches to zero for both protocols. Consequently, the optimization remains physically meaningful, as it aims to maximize the rate of symmetry restoration. This highlights a key conceptual difference between the two regimes: while imaginary-time optimization targets a unique ground state, real-time optimization focuses on accelerating the dynamical process of symmetry recovery.

These results demonstrate that the optimized configuration generates a trajectory with a significantly faster decay rate than the unoptimized perturbation. This confirms that our variational method successfully identifies a uniquely efficient dynamical path toward equilibrium within the variational landscape. Therefore, the principle of the Pontus-Mpemba effect, integrated with variational optimization, provides practical engineering guidance for fast state preparation.

\section{Discussions and Conclusion}
In this study, we introduce and investigate the concept of quantum Pontus-Mpemba effects, a counterintuitive phenomenon where a quantum system relaxes faster through a two-step evolution protocol compared to a direct quench. By employing both real-time and imaginary-time dynamics, we demonstrate that QPME emerges in systems with $U(1)$ symmetry, particularly when initialized in tilted ferromagnetic states at small angles. In real-time, the asymmetric evolution of the two-step protocol increases the thermal nature of the initial state, resulting in significantly accelerating thermalization. For imaginary-time evolution, the two-step protocol also accelerates convergence to the ground state for small tilt angles. We attribute this speedup to the broader charge distribution generated by the initial asymmetric evolution. Conversely, QPME is suppressed for large tilt angles or for antiferromagnetic initial states. Our variational analysis in Sec. IV further confirms that this acceleration is not limited to specific heuristically chosen Hamiltonians; rather, an optimized symmetry-breaking path exists that enhances relaxation efficiency. Notably, our results for system sizes up to $L=48$ show consistent QPME behavior, which provides supportive evidence of its potential robustness in the thermodynamic limit.

While of fundamental interest, the true strength of the QPME protocol lies in its practical applicability. Whereas QME compares the dynamics evolving from different initial states to the same target state, the QPME protocol relies on just a single initial state. This distinction is operationally significant: in many experimental settings, the central challenge is not to compare different starting conditions, but to identify an efficient dynamical path from a given initial state toward a final target state. The QPME protocol directly addresses this need by leveraging transient symmetry breaking to accelerate the relaxation process. Furthermore, the principles of QPME in imaginary time dynamics suggest a powerful new strategy for enhancing the performance of numerical algorithms. For tensor network methods like TEBD, a shorter evolution time translates to fewer computational steps, leading to faster ground state preparation. This speed-up is also particularly impactful for projection quantum Monte Carlo (PQMC) simulations. In many interesting systems, PQMC is plagued by the infamous sign problem, where computational complexity grows exponentially with the projection time. By achieving convergence in a shorter imaginary time, our protocol can substantially mitigate the sign problem \cite{hirsch1986monte,loh1990sign,zhang1997constrained,troyer2005computational,assaad2008world,kaul2013bridging,li2015solving,li2019sign,wan2022mitigating}, reducing the computational resources required and potentially extending the reach of PQMC to previously intractable problems.

Hence, our findings not only elucidate an intriguing non-equilibrium phenomenon but also provide a valuable strategy for accelerating quantum simulations and state preparation. Future work could extend these results to other symmetries, open quantum systems, or more complex interaction geometries.

\section*{ACKNOWLEDGMENTS}
The work is supported by the Ministry of Science and Technology  (Grant No. 2022YFA1403900), the Strategic Priority Research Program of the Chinese Academy of Sciences (Grant Nos. XDB28000000, XDB33000000), and the New Cornerstone Investigator Program. HY is supported by the International Young Scientist Fellowship of the Institute of Physics, Chinese Academy of Sciences (No.202407). SXZ acknowledges the support from the Innovation Program for Quantum Science and Technology (2024ZD0301700) and the National Natural Science Foundation of China (No. 12574546).

\section*{Data availability.}  Numerical data for this manuscript are publicly accessible in Ref. \cite{data-available}.

\label{LastBibItem}
\bibliography{reference}

\end{document}


\renewcommand{\thefigure}{S\arabic{figure}}
\setcounter{figure}{0}
\renewcommand{\theequation}{S\arabic{equation}}
\setcounter{equation}{0}
\renewcommand{\thesection}{\Roman{section}}
\setcounter{section}{0}
\setcounter{secnumdepth}{4}
\begin{center}
\textbf{\large Supplemental Material for ``Quantum Pontus-Mpemba effects in Real and Imaginary-time Dynamics''}
\end{center}

\author{Hui Yu}
\affiliation{Beijing National Laboratory for Condensed Matter Physics and Institute of Physics, Chinese Academy of Sciences, Beijing 100190, China}

\author{Jiangping Hu}
\email{jphu@iphy.ac.cn}

\affiliation{Beijing National Laboratory for Condensed Matter Physics and Institute of Physics, Chinese Academy of Sciences, Beijing 100190, China}
\affiliation{New Cornerstone Science Laboratory, Institute of Physics, Chinese Academy of Sciences, Beijing 100190, China}

\author{Shi-Xin Zhang}
\email{shixinzhang@iphy.ac.cn}

\affiliation{Beijing National Laboratory for Condensed Matter Physics and Institute of Physics, Chinese Academy of Sciences, Beijing 100190, China}
\maketitle

\section{The chaotic nature of the Hamiltonian $H$}
The Hamiltonian $H$ studied in the main text is defined as:
\begin{eqnarray}
H = & - \sum_{j=1}^{L}(\sigma_j^x \sigma_{j+1}^x + \gamma \sigma_j^y \sigma_{j+1}^y + \mu \sigma_j^z \sigma_{j+1}^z) \label{eq:Ham2}
+ \sum_{j=1}^{L} h_{j}\sigma_j^z.
\label{eq:Ham2}
\end{eqnarray}
The coupling strength along the $z$-direction is fixed at $\mu=-0.5$. Disorder is introduced through random fields $h_{j}$, each sampled independently and uniformly from the interval $[-W, W]$ with disorder strength $W=1$. 

To determine the chaotic nature of $H$, we compute the level spacing ratio $r$ \cite{oganesyan2007localization} within the half-filling sector when $\gamma=1$. The $n$-th level spacing ratio is defined as:
\begin{eqnarray}
r_{n}=\frac{min(\Delta_{n},\Delta_{n+1})}{max(\Delta_{n},\Delta_{n+1})}
\end{eqnarray}
where $\Delta_{n}=E_{n+1}-E_{n}$ and $E_{n}$ is the $n$-th eigenenergy in ascending order. The level spacing ratio $r$ is obtained by averaging $r_{n}$ over different energy levels $n$ and across an ensemble of disorder configurations. Our calculation shows $r\approx0.53$, which matches the theoretical value for the quantum chaotic regime and confirms that our system is in that phase.

\section{Similarity transformation of the Hamiltonian}
Previously, our initial states were taken as a tilted ferromagnetic state, defined by:
\begin{eqnarray}
    \vert \psi_{i} (\theta)\rangle =  e^{-i\frac{\theta}{2} \sum_{j} \sigma_{j}^{y}} \vert 000...0 \rangle = U_{\theta} \vert 000...0 \rangle
    \label{eq: tilted ferromagnetic}
\end{eqnarray}
where we define the unitary operator $U_{\theta} =e^{-i\frac{\theta}{2} \sum_{j} \sigma_{j}^{y}}$, and the Hamiltonian $H$ is given by:
\begin{eqnarray}
H = & - \sum_{j=1}^{L}(\sigma_j^x \sigma_{j+1}^x + \gamma \sigma_j^y \sigma_{j+1}^y + \mu \sigma_j^z \sigma_{j+1}^z) \label{eq:Ham2} 
+ \sum_{j=1}^{L} h_{j}\sigma_j^z. 
\label{eq:Ham2}
\end{eqnarray}
If we restrict the initial state to the ferromagnetic state $\vert 000...0 \rangle$, the dynamics is governed by a transformed Hamiltonian $H^{'}$ defined as:
\begin{eqnarray}
H^{'} = U_{\theta}^{\dagger}HU_{\theta}
\label{eq:Ham2}
\end{eqnarray}
To derive $H^{'}$, we use the following identities:
\begin{eqnarray}
e^{i\frac{\theta}{2} \sigma^{y}}\sigma^{x}e^{-i\frac{\theta}{2} \sigma^{y}} = \sigma^{x}\cos{\theta}-\sigma^{z}\sin{\theta}
\label{eq:x_rot}
\end{eqnarray}
\begin{eqnarray}
e^{i\frac{\theta}{2} \sigma^{y}}\sigma^{z}e^{-i\frac{\theta}{2} \sigma^{y}} = \sigma^{z}\cos{\theta}+\sigma^{x}\sin{\theta}
\label{eq:z_rot}
\end{eqnarray}
Applying these, we obtain the transformed nearest-neighbor terms:
\begin{eqnarray}
U_{\theta}^{\dagger}\sigma_{j}^{x}\sigma_{j+1}^{x}U_{\theta} = (\sigma_{j}^{x}\cos{\theta}-\sigma_{j}^{z}\sin{\theta})(\sigma_{j+1}^{x}\cos{\theta}-\sigma_{j+1}^{z}\sin{\theta})
\end{eqnarray}
\begin{eqnarray}
U_{\theta}^{\dagger}\sigma_{j}^{z}\sigma_{j+1}^{z}U_{\theta} = (\sigma_{j}^{z}\cos{\theta}+\sigma_{j}^{x}\sin{\theta})(\sigma_{j+1}^{z}\cos{\theta}+\sigma_{j+1}^{x}\sin{\theta})
\end{eqnarray}
The $\sigma^{z}$ term transforms as:
\begin{eqnarray}
U_{\theta}^{\dagger}\sigma_{j}^{z}U_{\theta} = \sigma_{j}^{z}\cos{\theta}+\sigma_{j}^{x}\sin{\theta}
\end{eqnarray}
Grouping and simplifying all terms, we arrive at the expression for $H^{'}$:
\begin{eqnarray}
H^{'} = & - \sum_{j=1}^{L} \big[ (\cos^{2}{\theta}+\mu\sin^{2}{\theta})\sigma_j^x \sigma_{j+1}^x + \gamma \sigma_j^y \sigma_{j+1}^y + (\sin^{2}{\theta}+\mu\cos^{2}{\theta}) \sigma_j^z \sigma_{j+1}^z \big] \label{eq:newHam} \\
& - \sum_{j=1}^{L} \big[\sigma_j^x \sigma_{j+1}^z+\sigma_j^z \sigma_{j+1}^x \big](\mu-1)\sin{\theta}\cos{\theta}+ \sum_{j=1}^{L} h_{j}(\sigma_{j}^{z}\cos{\theta}+\sigma_{j}^{x}\sin{\theta}). \notag
\label{eq:Ham2}
\end{eqnarray}
Therefore, a $U(1)$-symmetric evolution starting from a tilted ferromagnetic state is equivalent to a $U(1)$-asymmetric evolution starting from a ferromagnetic state. For initial states with a small tilt angle $\theta$ and anisotropy parameter $\gamma$ close to $1$, the symmetry-breaking terms proportional to $1-\gamma$ and $\sin{\theta}$ can be treated as perturbations. Consequently, the initial state has significant overlap with only a small number of eigenstates (scaling polynomially with system size). This mechanism, known as the Hilbert subspace imprint \cite{yu2025hilbert}, explains the observed slow thermalization for states with a small initial tilt.

\section{More numerical results for real-time entanglement asymmetry}

\begin{figure}[htbp]
\begin{center}
\includegraphics[scale=0.7]{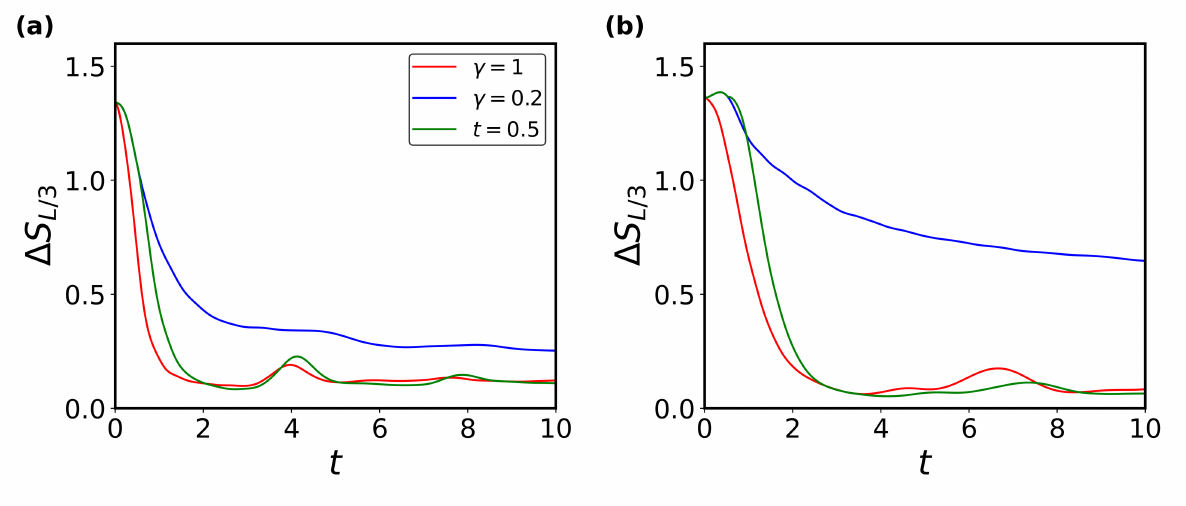}
\caption{Real-time dynamics of entanglement asymmetry $\Delta S_{L/3}$ for initial states with tilt angle $\theta=0.4\pi$. (a) Tilted ferromagnetic state and (b) tilted antiferromagnetic state. Evolution protocols compared: symmetric (red,$\gamma=1$), asymmetric (blue,$\gamma=0.2$), and two-step quench (green) with Hamiltonian switching from asymmetric to symmetric at $t=0.5$.}
\label{fig:EA_real_time_AF}
\end{center}
\end{figure}

At $\theta=0.4\pi$, no evidence of the pronounced quantum Pontus-Mpemba effect (QPME) is observed in either the tilted ferromagnetic or tilted antiferromagnetic initial states, as illustrated in Fig.~\ref{fig:EA_real_time_AF}(a) and (b). For tilted ferromagnetic states, the absence of QPME is attributed to the large tilt angle, which promotes a more uniform distribution of the initial state across charge sectors. This increased thermalization effectively diminishes the influence of transient asymmetry evolution. Similarly, for tilted antiferromagnetic states, the QPME is not observed because the untilted Néel state already possesses a highly thermal character due to its maximal Hilbert space dimension. In this instance, the transient symmetry-breaking evolution is insufficient to induce a noticeable QPME.

\section{Dependence of QPME on switching time and tilt angle}

\begin{figure}[ht]
\centering
\includegraphics[width=0.48\textwidth, keepaspectratio]{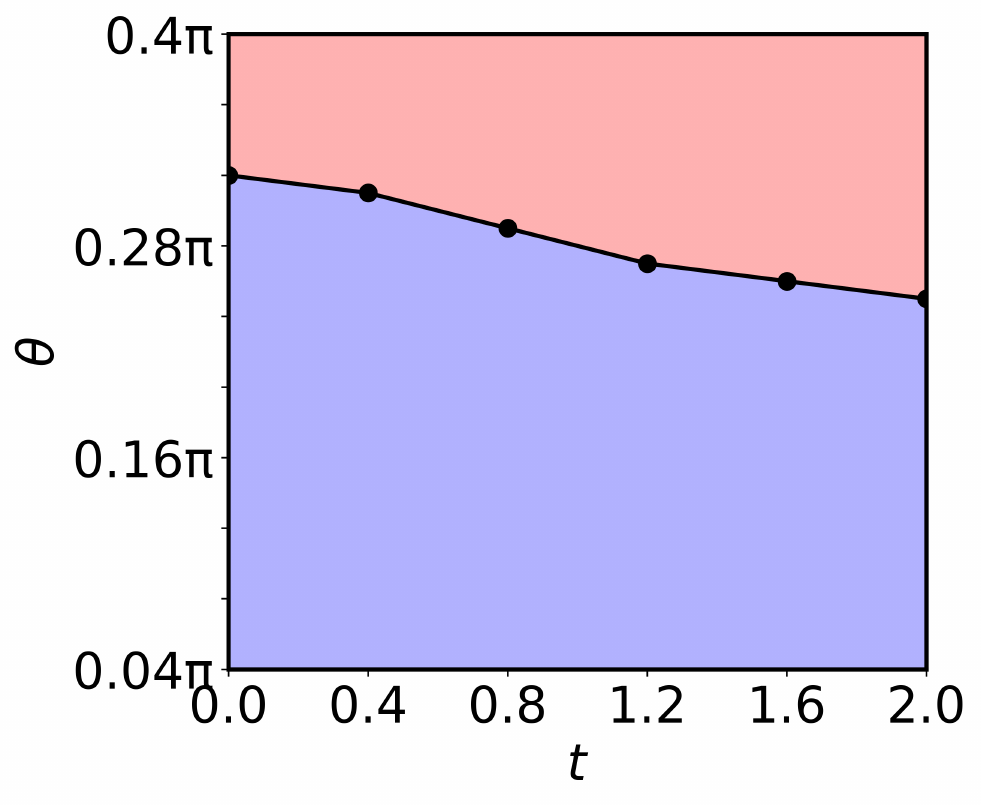}
\caption{The emergence of the QPME in a ferromagnetic state of size $L=12$ shows its dependence on the tilt angle $\theta$ and the switching time $t$. The QPME is identified by comparing the relaxation of the entanglement asymmetry following two distinct protocols: a 2-step protocol, where the system first evolves under the asymmetric Hamiltonian $H_{asym}$ with $\gamma=0.6$, and a direct quench. Regions in blue indicate parameters for which the 2-step protocol's dynamics cross below those of the direct quench during relaxation, signaling the presence of the QPME. In contrast, red regions signify its absence, as the 2-step curve remains above the direct quench curve throughout the relaxation process.}
\label{fig:qpme_scan}
\end{figure}

We investigate the emergence of the QPME across the parameter space of tilt angle $\theta$ and switching time $t$ by comparing the relaxation dynamics of the entanglement asymmetry for a ferromagnetic state following the 2-step protocol versus a direct quench. As depicted in Fig.~\ref{fig:qpme_scan}, the blue and red regions correspond to the presence and absence of the QPME, respectively. We observe that the QPME is prominent for weak symmetry breaking (small $\theta$), consistent with the theory of HSI. The specific parameters ($\theta=0.2\pi$ and $t=0.5$) from Fig.1(a) of the main text are confirmed to lie within the QPME regime, validating our earlier analysis.

\section{The charge sector of the ground state in the Hamiltonian $H_{sym}$}

\begin{figure}[htbp]
\begin{center}
\includegraphics[scale=0.52]{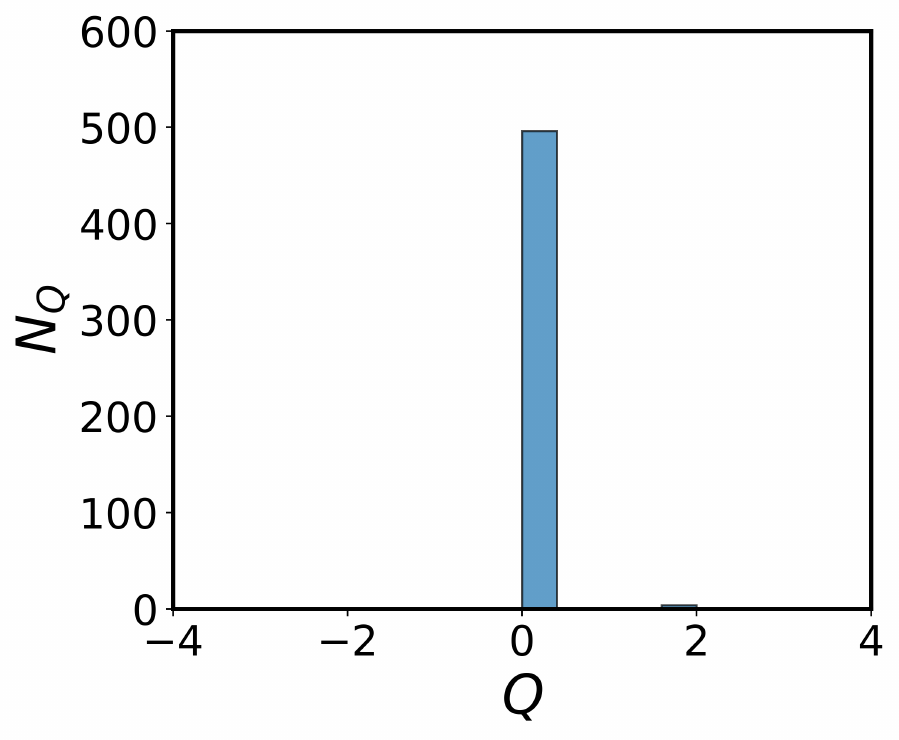}
\caption{$N_{Q}$ counts the number of ground states of $H_{sym}$ in charge sector $Q$, computed using $500$ independent disorder samples.}
\label{fig:GS_Qsector}
\end{center}
\end{figure}

Fig.~\ref{fig:GS_Qsector} shows the charge sector distribution of the ground state for the $U(1)$-symmetric Hamiltonian $H_{sym}$ calculated across $500$ disorder realizations. The ground state overwhelmingly occupies the $Q=0$ sector, with only an extremely small fraction of realizations exhibiting a $Q=2$ ground state.

\section{Energy relaxation dynamics of tilted ferromagnetic states under $H_{sym}$}

\begin{figure}[htbp]
\begin{center}
\includegraphics[scale=0.52]{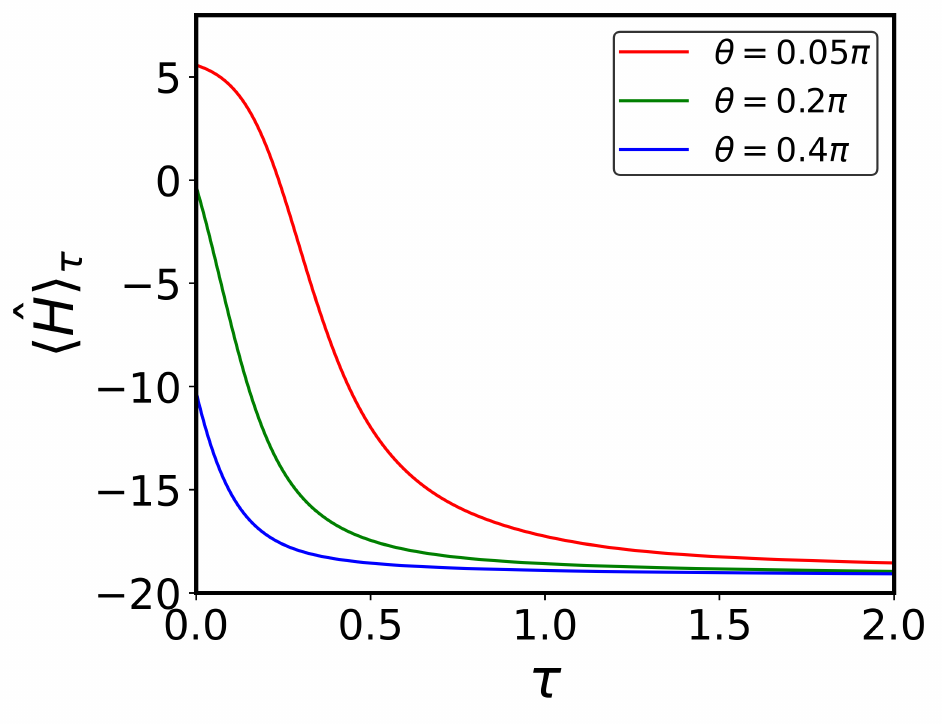}
\caption{The imaginary-time evolution of the energy expectation value $\langle \hat{H} \rangle_{\tau}$ is presented for various tilted ferromagnetic states. All dynamics are govern by the symmetric Hamiltonian $H_{sym}$.}
\label{fig:GS_difftheta_F}
\end{center}
\end{figure}

Fig.~\ref{fig:GS_difftheta_F} shows the imaginary-time evolution of $\langle \hat{H} \rangle_{\tau}$ for tilted ferromagnetic states. The observed acceleration in energy relaxation with increasing tilt angle $\theta$ is driven by the broader initial charge distributions, which provide more efficient pathways to the ground state.

\section{Comparing charge sector dynamics of tilted ferromagnetic state under two protocols}

\begin{figure}[htbp]
\begin{center}
\includegraphics[scale=0.7]{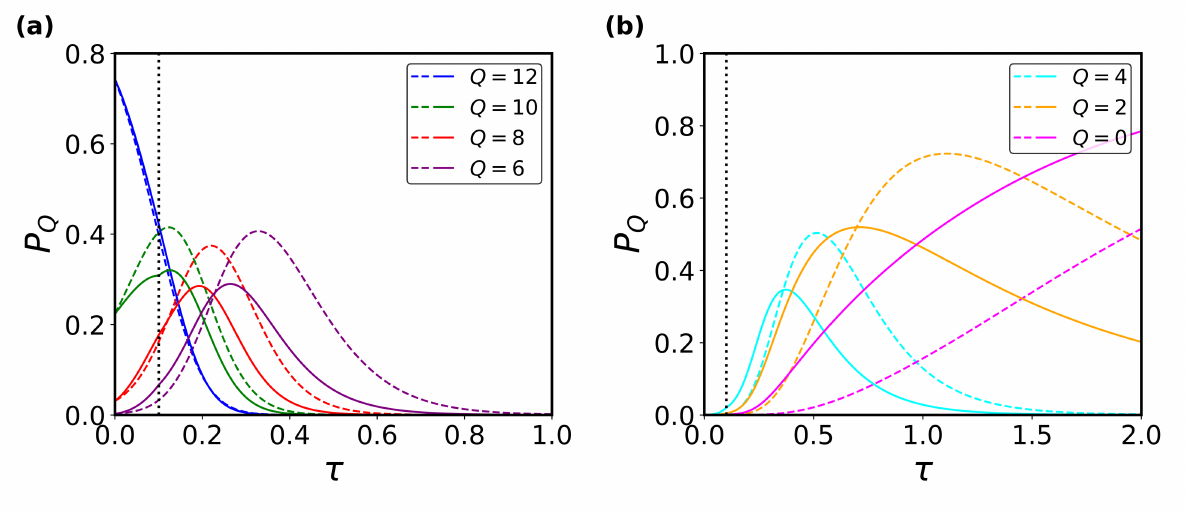}
\caption{Imaginary-time evolution of the probability distribution, $P_{Q}$, across sectors $Q$ for a tilted ferromagnetic initial state ($\theta=0.1\pi$). Results are shown for two protocols: direct quench (dashed lines) and a two-step protocol (solid lines), where the system first evolves under the asymmetric Hamiltonian $H_{asym}$ ($\gamma=0.2$) before switching to $H_{sym}$ at time $\tau=0.1$ (marked by the vertical dotted line). Populations for negative $Q$ are negligible and are not shown.}
\label{fig:ChargeS_F}
\end{center}
\end{figure}

\begin{figure}[htbp]
\begin{center}
\includegraphics[scale=0.27]{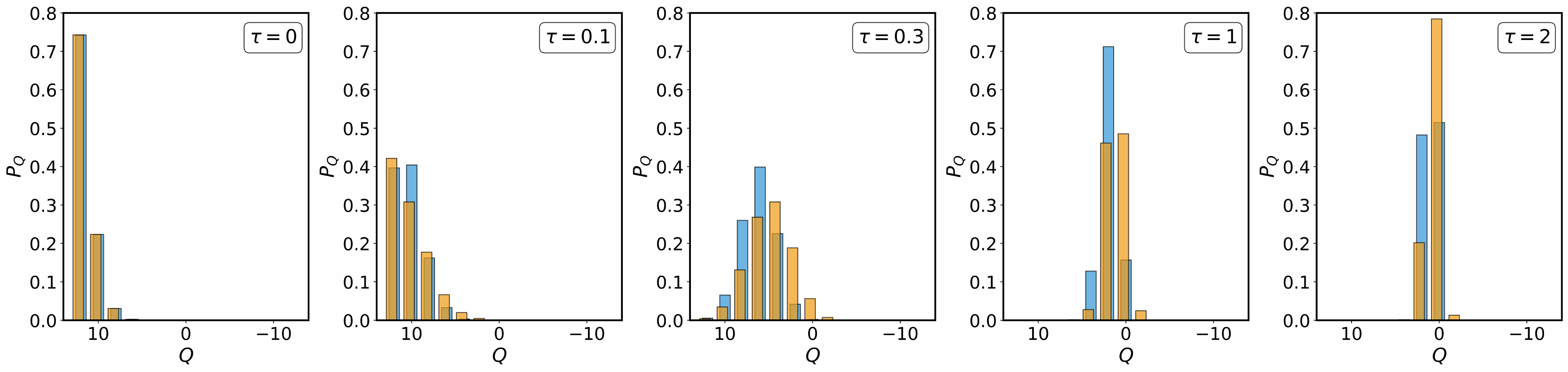}
\caption{Imaginary-time evolution of the probability distribution, $P_{Q}$ across charge sectors $Q$ (ranging from $-12$ to $12$ in increments of 2) for a tilted ferromagnetic initial state ($\theta=0.1\pi$), comparing two protocols: direct quench (blue) and a two-step protocol (orange) involving initial evolution under the asymmetric Hamiltonian $H_{asym}$ ($\gamma=0.2$). The columns, from left to right, show the probability distributions at evolution times $\tau=0$, $0.1$, $0.3$, $1$, and $2$.}
\label{fig:ChargeS_slice_F}
\end{center}
\end{figure}

Fig.~\ref{fig:ChargeS_F} displays the probability distribution $P_Q$ across various charge sectors $Q$ for the evolved state $|\psi(\tau)\rangle$ for initial tilted ferromagnetic states. Here, $P_{Q}$ at time $\tau$ is defined as $\sum_{q=Q}|\langle \psi(\tau)|\psi_{q}\rangle|^{2}$, where $|\psi_{q}\rangle$ represents the basis wave function corresponding to charge $q$. States evolved under transient asymmetry evolution exhibit a broader charge distribution compared to those undergoing direct quench. Furthermore, sectors with $Q\neq0$ decay much more rapidly in the two-step protocol, resulting in faster convergence to the ground state. The temporal evolution of the probability distribution across all charge sectors is shown at selected time slices in Fig.~\ref{fig:ChargeS_slice_F}, revealing a gradual shift from $Q=12$ toward $Q=0$ as $\tau$ increases, with the two-step protocol demonstrating significantly accelerated dynamics.

\section{Eigenstate overlap probability distribution of tilted ferromagnetic initial states}.

\begin{figure}[htbp]
\begin{center}
\includegraphics[scale=0.52]{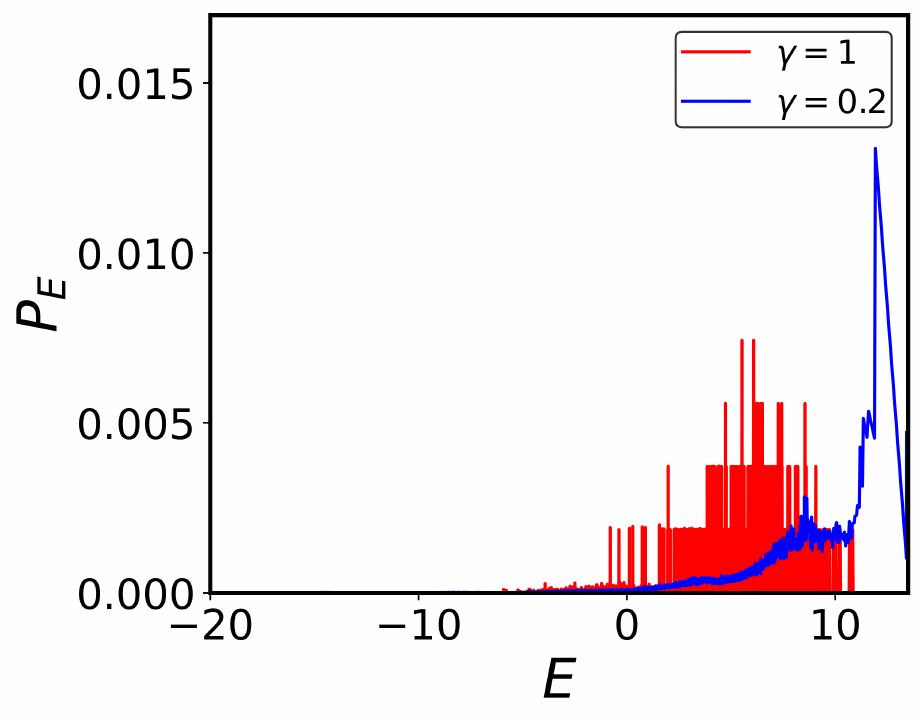}
\caption{Overlap probability distribution, $P_{E}$, between tilted ferromagnetic states ($\theta=0.05\pi$) and eigenstates of energy $E$. Red curves denote overlaps with eigenstates of the symmetric Hamiltonian $H_{sym}$, while blue curves represent overlaps with eigenstates of the asymmetric Hamiltonian $H_{asym}$ ($\gamma=0.2$).}
\label{fig:GS_eigenoverlap}
\end{center}
\end{figure}

As shown in Fig.~\ref{fig:GS_eigenoverlap}, the eigenstate overlap distributions differ significantly between $H_{sym}$ and $H_{asym}$. For the $U(1)$-symmetric Hamiltonian $H_{sym}$, the overlap is predominantly concentrated in the energy range from $0$ to $10$. In contrast, for the $U(1)$-asymmetric Hamiltonian $H_{asym}$, most of the spectral weight lies at energies above $10$. This key distinction explains why energy relaxation occurs faster in $H_{asym}$ compared to $H_{sym}$, as discussed in the main text.

\section{Energy dynamics comparison across different switching times}

\begin{figure}[htbp]
\begin{center}
\includegraphics[scale=0.85]{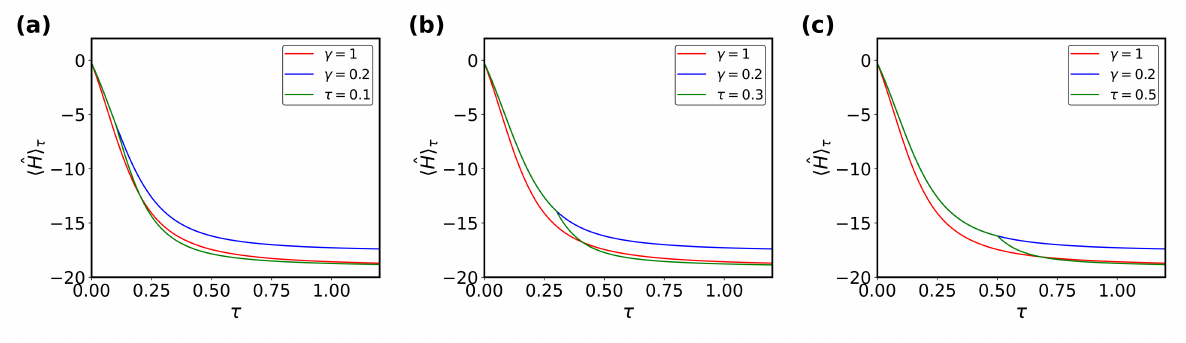}
\caption{The imaginary-time dependence of $\langle \hat{H} \rangle_{\tau}$ is shown for tilted ferromagnetic states with $\theta=0.2\pi$ at different switching times. The red and blue curves correspond to symmetric and asymmetric (with $\gamma=0.2$) evolution, respectively, while the green curves represent the two-step protocol, where the Hamiltonian switches from asymmetric to symmetric at (a) $\tau=0.1$, (b) $\tau=0.3$, and (c) $\tau=0.5$.}
\label{fig:QPME_time_F}
\end{center}
\end{figure}

\begin{figure}[htbp]
\begin{center}
\includegraphics[scale=0.85]{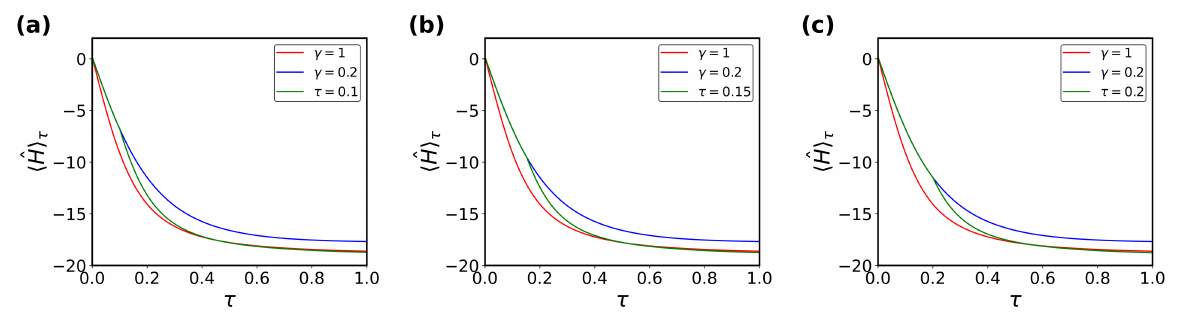}
\caption{The imaginary-time dependence of $\langle \hat{H} \rangle_{\tau}$ is shown for tilted antiferromagnetic states with $\theta=0.2\pi$ at different switching times. The red and blue curves correspond to symmetric and asymmetric (with $\gamma=0.2$) evolution, respectively, while the green curves represent the two-step protocol, where the Hamiltonian switches from asymmetric to symmetric at (a) $\tau=0.1$, (b) $\tau=0.15$, and (c) $\tau=0.2$.}
\label{fig:QPME_time_AF}
\end{center}
\end{figure}

In Fig.~\ref{fig:QPME_time_F}, we analyze the energy relaxation dynamics under the two-step protocol for varying switching times $\tau_{switch}$. The QPME exhibits a strong dependence on $\tau_{switch}$. Here, we denote $\tau_{relax}$ as the relaxation timescale of the purely asymmetric Hamiltonian evolution (blue curves). For early switching ($\tau_{switch} \ll \tau_{relax}$), QPME significantly accelerates relaxation. As $\tau_{switch}$ approaches $\tau_{relax}$, the QPME weakens progressively. When $\tau_{switch} \gg \tau_{relax}$, QPME vanishes completely, rendering the protocol ineffective. To preserve pronounced QPME, the Hamiltonian must transition from asymmetric to symmetric before the asymmetric evolution stabilizes. Notably, for tilted antiferromagnetic states, the QPME remains absent across all switching times, as shown in Fig.~\ref{fig:QPME_time_AF}.

\section{Comparison of trace distance dynamics in 2-step and direct quench protocols}

\begin{figure}[ht]
\centering
\includegraphics[width=0.8\textwidth, keepaspectratio]{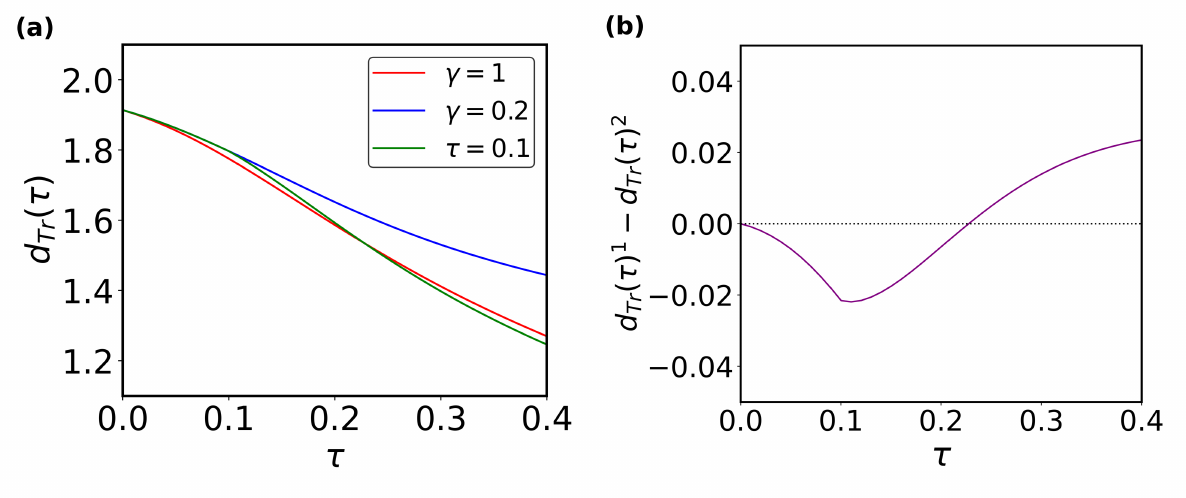}
\caption{(a) The imaginary-time evolution of the trace distance $d_{Tr}(\tau)$ is shown for tilted ferromagnetic states with $\theta=0.1\pi$ at system sizes $L=12$. The red and blue curves correspond to symmetric and asymmetric evolution, respectively, with the asymmetric evolution governed by Hamiltonian $H_{asym}$ with $\gamma=0.2$. The green curve illustrates the two-step protocol with a switch time set to $\tau=0.1$.  (b) The trace distance difference between the two protocols, defined as $d_{Tr}(\tau)^{1}-d_{Tr}(\tau)^{2}$, is plotted as a function of $\tau$, where $d_{Tr}(\tau)^{1}$ and $d_{Tr}(\tau)^{2}$ denote the trace distances for the direct quench and two-step protocol, respectively.}
\label{fig:trace_d}
\end{figure}

In the main text, the QPME is characterized using entanglement asymmetry and energy. To investigate whether QPME is a more general phenomenon manifesting in other physical quantities, we study the dynamics of the trace distance. The trace distance provides another measure of distinguishability between two quantum states. We focus on the evolution of a subsystem $A$. Its trace distance at imaginary time $\tau$ is defined as:
\begin{equation}
d_{tr}(\tau) = \Tr(\sqrt{M(\tau)^{\dagger}M(\tau)})   
\end{equation}
where $M(\tau)=\rho_{A}(\tau)-\rho_{A, final}(\tau\rightarrow\infty)$. $\rho_{A}(\tau)$ is the reduced density matrix of $A$ at time $\tau$, and $\rho_{A, final}$ is its target state in the ground state of $H_{sym}$. As shown in Fig.~\ref{fig:trace_d}(a), the trace distance for the two-step protocol (green curve) crosses below that of the direct quench (red curve) around $\tau=0.25$, clearly exhibiting the QPME. This crossing is further corroborated by the sign change in the trace distance difference between the two protocols, plotted in Fig.~\ref{fig:trace_d}(b). The manifestation of the QPME in this state-based metric confirms that the effect is robust and not an artifact of a specific observable. 

\section{Variational optimization of the dynamical path in real-time}
\begin{figure}[htbp]
\begin{center}
\includegraphics[width=0.8\textwidth, keepaspectratio]{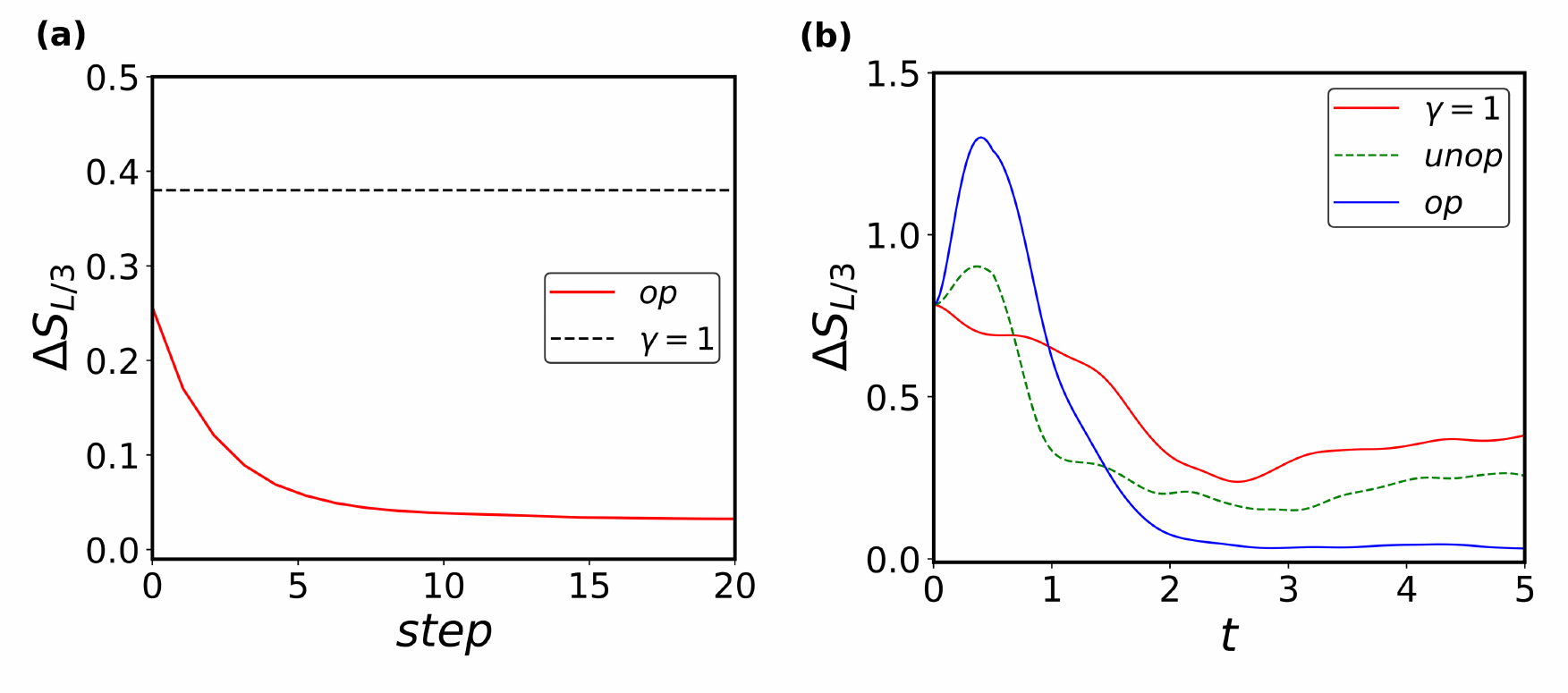}
\caption{Variational identification of the optimal relaxation path for $L=12$ and tilted ferromagentic state with $\theta=0.2\pi$. (a) Convergence of the loss function (the entanglement asymmetry at $t_{total}=5$) over optimization steps. The switching time is fixed at $t=0.5$. The black dashed line marks the reference value for a system governed solely by $H_{sym}$. (b) Real-time evolution of entanglement asymmetry. The red line shows the standard direct quench under $H_{sym}$. The two-step protocols involve an initial evolution under $H_{opt}$ followed by $H_{sym}$. The blue line utilizes the optimized parameters $\{\gamma_{j},h_{j}\}$, while the green dashed line uses unoptimized parameters.}
\label{fig:optimal_real}
\end{center}
\end{figure}

We perform a variational optimization to identify the optimal relaxation path in real-time dynamics. We parameterize the transient Hamiltonian $H_{opt}$ using the same form as in the main text:
\begin{eqnarray}
H_{opt}(\{\gamma_{j}\},\{h_{j}\}) = & - \sum_{j=1}^{L}(\sigma_j^x \sigma_{j+1}^x  + \mu \sigma_j^z \sigma_{j+1}^z) \notag \\
& - \sum_{j=1}^{L}\gamma_{j}\sigma_j^y \sigma_{j+1}^y + \sum_{j=1}^{L} h_{j}\sigma_j^z.
\end{eqnarray}
The variational parameters consist of bond-dependent anisotropies $\gamma_{j} \in[0,1]$, which control the local strength of the symmetry breaking, and site-dependent longitudinal fields $h_{j} \in [-1,1]$. 

The objective is to minimize the entanglement asymmetry at a final time $t_{total}$ with respect to the target symmetric Hamiltonian $H_{sym}$ (where $\gamma_{j}=1$ for all $j$). The state $|\psi(t_{total})\rangle$ is obtained by evolving the initial state first under $H_{opt}(\{\gamma_{j}\},\{h_{j}\}) $ for a duration $t$, and subsequently under $H_{sym}$ for the remainder of the time. In the optimization procedure, we enforce physical constraints on the variational parameters to ensure that $\gamma_{j} \in[0,1]$ and $h_{j} \in [-1,1]$ throughout the search. To compare with the unoptimized results from Fig.~1(a) of the main text, we perform this optimization for the real-time dynamics of a tilted ferromagnetic state ($\theta=0.2\pi$) with a system size $L=12$. We set the total evolution time $t_{total}=5$ and the switching time $t=0.5$. The optimizer is initialized with uniform couplings $\gamma_{j}=0.6$ for all $j$ and random longitudinal fields, using a learning rate of $0.4$.

As shown in Fig.~\ref{fig:optimal_real}(a), the entanglement asymmetry decreases monotonically as the optimization progresses and saturates after approximately 10 steps. To verify that this configuration indeed yields a faster relaxation rate, we analyze the time dependence of the entanglement asymmetry in Fig.~\ref{fig:optimal_real}(b). When employing the optimized parameters for $H_{opt}$ (blue line), the system exhibits enhanced QMPE behavior, with the entanglement asymmetry remaining consistently lower than unoptimized curve (green dashed line) for $t>1.5$. However, one must interpret the notion of an optimal path with caution in this context. Unlike imaginary-time evolution where the target is a unique ground state, real-time dynamics leads to thermalization at a steady state determined by the system's energy density. Since different transient Hamiltonians $H_{opt}$ inject different amounts of energy relative to $H_{sym}$, the unoptimized and optimized protocols technically relax toward thermal ensembles with different effective temperatures. Consequently, the optimization here specifically maximizes the rate of symmetry restoration, rather than convergence to a unique, fixed target state.

\bibliography{reference.bib}